\pdfoutput=1
\documentclass[%
 aip,
 jmp,%
 amsmath,amssymb,
 reprint,%
]{revtex4-1}
\usepackage{graphicx}
\usepackage{dcolumn}
\usepackage{amsmath}
\usepackage{bm}
\usepackage{textcomp}
\usepackage{pbox}
\usepackage{array,booktabs}
\usepackage{amsmath}
\usepackage{braket}

\begin{document}

\preprint{AIP/123-QED}

\title[]{Visualizing dispersive features in 2D image via minimum gradient method}

\author{Yu He}
\affiliation{
SIMES, SLAC National Accelerator Laboratory, Menlo Park, California 94025
}
\affiliation{
Department of Applied Physics, Stanford University, Stanford, California 94305
}
\author{Yan Wang}
\affiliation{
Microsoft, Redmond, WA 98052
}
\author{Zhi-Xun Shen}
\affiliation{
SIMES, SLAC National Accelerator Laboratory, Menlo Park, California 94025
}
\affiliation{
Department of Applied Physics, Stanford University, Stanford, California 94305
}
\email{zxshen@stanford.edu}

\date{\today}

\begin{abstract}
We developed a minimum gradient based method to track ridge features in 2D image plot, which is a typical data representation in many momentum resolved spectroscopy experiments. Through both analytic formulation and numerical simulation, we compare this new method with existing DC (distribution curve) based and higher order derivative based analyses. We find that the new method has good noise resilience and enhanced contrast especially for weak intensity features, meanwhile preserves the quantitative local maxima information from the raw image. An algorithm is proposed to extract 1D ridge dispersion from the 2D image plot, whose quantitative application to angle-resolved photoemission spectroscopy measurements on high temperature superconductors is demonstrated.
\end{abstract}

\maketitle


\section{\label{sec:level1}INTRODUCTION}
2D image plot is one of the most commonly used data representation formats in many physical researches, widely utilized in techniques like scanning tunneling microscopy (STM)\cite{Vang2008}, microwave impendence microscopy (MIM)\cite{Keji2011} and inelastic neutron scattering spectroscopy (INS)\cite{Laura2013}. With the rapid advancement in 2D detector technologies, 2D image plot becomes the native data acquisition unit for many energy-momentum resolved techniques, including resonant inelastic X-ray scattering (RIXS)\cite{huang2016}, angle-resolved photoemission spectroscopy (ARPES)\cite{Junfeng2016} and time-of-flight (TOF) based spectroscopy\cite{jozwiak2016}. For many quantitative purposes, the extrema (ridges or valleys) and the widths of the 2D data usually constitute the two most important pieces of information.

For example, in a typical energy-momentum ($\omega-k$) map from ARPES measurement, the spectral intensity can be described as $I_0(k,\omega)$ in Eq.~(\ref{eq:1})~-~(\ref{eq:4}):
\begin{equation}
\begin{split}
I_0(k,\omega)~\sim~ & \{A(k,\omega) \times FD(\omega) \times |M(k)|^2 + g_n^{in}(k,\omega)\}\\
& \otimes R(k,\omega) + g_n^{ex}(k,\omega)
\label{eq:1}
\end{split}
\end{equation}

\begin{equation}
A(k,\omega) = -\frac{1}{\pi} \frac{Im\Sigma(k,\omega)}{(\omega - \epsilon(k) - Re\Sigma(k,\omega))^2 + (Im\Sigma(k,\omega))^2}
\tag{2}\label{eq:2}
\end{equation}

\begin{equation}
FD(\omega) = \frac{1}{e^{\frac{\omega}{k_BT_e}} + 1}
\tag{3}\label{eq:3}
\end{equation}

\begin{equation}
M(k) = \Bra{\phi_f^{k}}-\frac{e}{mc}{A}\cdot{p}\Ket{\phi_i^{k}}
\tag{4}\label{eq:4}
\end{equation}

Here $A(k,\omega)$ is the single particle spectral function that contains the critical information of the electron bare band dispersion $\epsilon(k)$ and self energy $\Sigma(k,\omega)$ in Eq.~(\ref{eq:2}). $FD(\omega)$ is the electronic Fermi-Drac distribution modulating the intensity's energy distribution (Eq.~(\ref{eq:3})). $M(k)$ is the single electron dipole transition matrix element modulating the intensity's momentum distribution. $R(k,\omega)$ is the instrument energy-momentum resolution function. $g_n^{in}(k,\omega)$ and $g_n^{ex}(k,\omega)$ represent the intrinsic (sample surface vibration, phosphor screen roughness) and extrinsic (CCD sensor, electrical circuit) experimental noise respectively.

 \begin{figure*}[t]
 \includegraphics[width=0.9\textwidth]{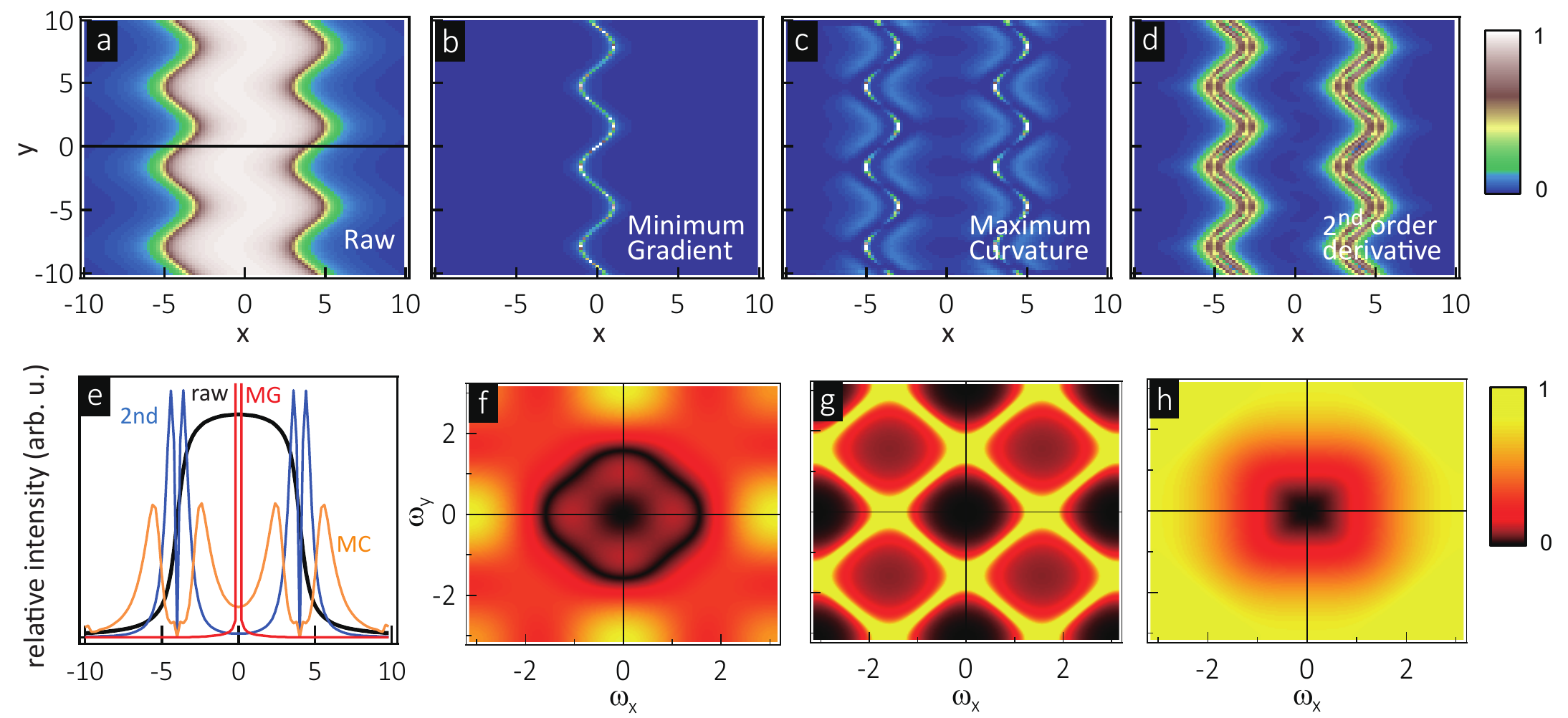}
 \caption{\label{Fig0} (a) Simulated wiggly dispersion with plateau horizontal lineshape. The functional form is $z = \text{atan}(2(x-\text{sin}(y))+8) - \text{atan}(2(x-\text{sin}(y))-8)$. (b)(c)(d) Dispersion extracted with minimum gradient (MG) method, maximum curvature method (MC) and second derivative method (square sum of both x and y derivatives). (e) Line cut at $y = 0$ for (a)-(d). (f)(g)(h) Frequency response of the MG, MC and second derivative method. 
 }
 \end{figure*}

\begin{table*}[t]
\resizebox{\textwidth}{!}{%
\centering
\setlength\extrarowheight{15pt}
\begin{tabular}{cccccccc}

\hline
 & \pbox{20cm}{Orders of\\derivative} & \pbox{20cm}{peak FWHM\\Lorenztian/Gaussian} & \pbox{20mm} {peak position\\selection criteria} & \pbox{20cm} {noise\\strength} &  \pbox{20cm} {data acquisition time\\to compensate noise} & \pbox{20cm}{good for \\steep/flat\\dispersion} \\ \hline

EDC fitting\cite{armitage2003}  & 0 & 1/1 & $f^{(1)}(\omega)=0$ & $\sigma$ & 1 & flat \\
MDC fitting\cite{armitage2003} & 0 & 1/1 & $f^{(1)}(\omega)=0$ & $\sigma$ & 1 & steep \\
minimum gradient & 1 & \pbox{20cm}{0/0 (no noise)\\0.012/0.018 (3\% noise)} & $f^{(1)}(\omega)=0$ & $\frac{1.26\sigma}{\Delta \omega}$ & $\frac{1.59}{\Delta \omega}$ & both \\
maximum curvature\cite{Ding2011} & 2 & 0.093/0.253 & \pbox{20cm}{$f^{(1)}(\omega)=0$ and\\$f^{(3)}(\omega)=0$} & \textgreater~$\big(\frac{2\sigma}{\Delta \omega}\big)^2$ & \textgreater~$\frac{16 \sigma}{(\Delta \omega)^2}$ & both \\
$2^{nd}$ derivative\cite{yi2011} & 2 & 0.327/0.532 & $f^{(3)}(\omega)=0$ & $\big(\frac{2\sigma}{\Delta \omega}\big)^2$ & $\frac{16 \sigma}{(\Delta \omega)^2}$ & either \\ 
\hline

\end{tabular}}
\caption{Comparison of major techniques to visualize and extract 1D dispersion from 2D image plot data.}
\end{table*}

In order to extract the band dispersion and self energy, three techniques have been most frequently used - momentum/energy distribution curve (MDC/EDC) fitting method, second derivative method\cite{sato2001} and maximum curvature method\cite{Ding2011}.

MDC/EDC fitting first slices the 2D image into an array of 1D (constant energy or momentum) curves, then based on assumed self energy functional form (Eq.~(\ref{eq:2})), the peak position and peak width can be derived from 1D curve fitting. This approach, with properly chosen self energy and background, can yield physical quantities such as $Im\Sigma(k,\omega)$ and $(\epsilon(k) + Re\Sigma(k,\omega))$ and is noise resilient. However, EDC fitting is vulnerable to intensity modulation along $\omega$ axis, including that from $FD(\omega)$, rapidly dispersing $\epsilon(k)$, energetically close neighboring bands and the often unknown (thus model dependent) functional form of the self energy $\Sigma(k,\omega)$. On the other hand, while MDC fitting enjoys better defined Lorentzian functional form, its quantitative performance suffers from additional intensity modulation along $k$ axis, usually resulted from $|M(k)|^2$ and slowly dispersing $\epsilon(k)$ near band top/bottom.

To qualitatively sharpen the image plot, higher order derivative based methods have been employed to provide an enhanced visual guide. Second derivative method applies two consecutive derivatives of the spectral intensity along either $\omega$ axis (for slowly dispersing flat bands) or $k$ axis (for rapidly dispersing vertical bands). After this process, single peak feature in intensity plot is usually turned into one enhanced central valley with two weaker satellite peaks; and neighboring peak features are usually better separated due to the boosted sharpness. The peak can be sharpened to 53\% of the original width for peaks with Gaussian profiles, and 33\% for those with Lorentzian profiles. However, this method bears two significant drawbacks - sensitivity to noise corruption and altered peak position. First, differentiation serves as an effective high pass filter (Fig.~\ref{Fig0}(h)), therefore second order derivative method usually generates much noisier image. Second, extrema on a second order derivative image don't always reproduce the peak positions in the raw spectrum, because most functions don't simultaneously satisfy $I_0^{(3)}(\omega)=0$ and $I_0^{(1)}(\omega)=0$ on the same domain of definition (Fig.~\ref{Fig0}(e)). Moreover, in order to quench the noise in second derivative images, excessive smoothing is often practiced. This further introduces additional broadening of the spectrum and uncertainty of peak shifts due to noise, not to mention that the smoothing algorithm itself also has many different versions.

Recently, a maximum curvature method is proposed to tackle the aforementioned difficulties.\cite{Ding2011} This method computes the intensity function's local curvature based on either its 1D distribution curve curvature or a combination of 2D local curvatures. It successfully achieves even narrower FWHM with respect to the raw data (reduction down to 25\% for Gaussian profile and 9\% for Lorenztian profile). Still, due to the higher order derivative's nature of this method, both exacerbated noise corruption and peak shifting plight remain. Another vulnerability is that it tends to yield false detection at the edges of intensity plateau, instead of the actual intensity maxima (Fig.~\ref{Fig0}(c)(e)).

The different signal enhancing approaches can also be compared in the image's frequency domain (Fig.~\ref{Fig0}(f)-(h)). Typically a ~\lq feature\rq~ is identified if it is sufficiently separated from both the low frequency background and the high frequency noise. MDC/EDC fitting utilizes \emph{a-priori} information about the feature (functional form of the signal) to directly extract the feature via fitting. Higher order derivatives enhances the higher frequency component of the image (edges) and crop out low frequency (background) components. Post-processing techniques such as smoothing, binning and Gaussian blur (convolution) act as the low frequency filter to dampen excessive high frequency component, but they also downshift the signal frequency to that of the slow-varying background.

To improve the noise resilience and preserve the quantitative analyzability of the raw data, we propose a first order approach based on the minimum gradient distribution of the 2D function. Since the maxima of $A(k,\omega)$ are achieved by zeroing the $(\omega - \epsilon(k) - Re\Sigma(k,\omega))^2$ term on the denominator, this minimum gradient method captures the peak position thus the renormalized dispersion $\epsilon(k)+Re\Sigma(k,\omega)$ by definition. A comparison among the above methods are listed in Table I, grouped by effective orders of derivatives. Fig.~\ref{Fig0} also compares the latter three methods with a simulated 2D spectra (no noise corruption) in both the image primary domain and the frequency domain.

In this work, we describe the principles of this minimum gradient algorithm and compare the results with existing methods. Then the application to cuprate superconductor and multiband iron-based superconductor systems are demonstrated. We also propose a minimal model to extract 1D ridge features from the renormalized 2D minimum gradient map, and at last discuss its parameter stability.


\section{\label{sec:level1}MINIMUM GRADIENT METHOD}
\subsection{\label{sec:level2}The Algorithm and noise analysis}

For a second order differentiable bi-variable function $I_0(k,\omega)$, its local extremum $(k_0,\omega_0)$ must satisfy:
\begin{equation}
\partial_kI_0(k,\omega)|_{k_0} = \partial_\omega I_0(k,\omega)|_{\omega_0} = 0
\tag{5}\label{eq:5}
\end{equation}
Equivalently, the modulus of its gradient vector $G_{2D} = (\partial_kI_0,\partial_\omega I_0)$ reaches zero at all of its maxima, minima and saddle points. Therefore the 2D map of $||G_{2D}||^{-1}$ should diverge at and only at the local extrema of $I_0(k,\omega)$.

However, real experiments are always corrupted by noise $g_n$. It should be noted that for low photon/electron count experiments, Poisson distribution should be used to describe the noise instead. From Eqn.~(\ref{eq:1}) the total noise contribution can be expressed as:
\begin{equation}
\begin{split}
g_n ~\sim~ g_n^{in}(\sigma_{in}) \otimes R(\sigma_r) + g_n^{ex}(\sigma_{ex})
\label{eq:6}
\end{split}
\tag{6}
\end{equation}
where for simplicity we model both noise and resolution function with Gaussian profile. The intrinsic noise standard deviation $\sigma_{in}$ is often proportional to the incident light intensity, and $\sigma_{ex}$ for extrinsic noise is determined by the instrument setup. On the other hand, the instrument resolution convolution $g_n^{in}(\sigma_{in}) \otimes R(\sigma_r)$ acts as a Gaussian low pass filter to the intrinsic noise, which quenches the intrinsic noise to $\sigma_{in}/(2\sqrt{\pi}\sigma_r)$. Therefore the total noise from Eqn.~({\ref{eq:6}) can be simplified to a single gaussian with combined standard deviation of:
\begin{equation}
\begin{split}
\sigma = \sigma_{tot} = \frac{\sigma_{in}}{2\sqrt{\pi}\sigma_r} + \sigma_{ex}
\label{eq:7}
\end{split}
\tag{7}
\end{equation}

Upon difference quotient with the step size of $\delta$ along either axis, $g_n(\sigma)$ propagates as below if we assume energy-momentum $(k,\omega)$ independent 2D noise:
\begin{equation}
\begin{split}
&\partial_kg_n(\sigma) \sim \frac{g_n(\sigma)|_{(k+\delta,\omega)} - g_n(\sigma)|_{(k,\omega)}}{\delta} \sim g_n\big(\frac{2}{\delta}\sigma\big)\\
&...\\
&\partial_k^{(m)}g_n(\sigma) \sim \partial_\omega^{(m)}g_n(\sigma) \sim g_n\big(\frac{2^m}{\delta^m}\sigma\big)
\label{eq:8}
\end{split}
\tag{8}
\end{equation}
where $\partial_k^m$ stands for $m^{th}$ derivative along k direction. It is obvious that $m^{th}$ order differentiation acts as a high pass filter, boosting the noise standard deviation by $\big(\frac{2}{\delta}\big)^m$ times.

 \begin{figure*}
 \includegraphics[width=0.9\textwidth]{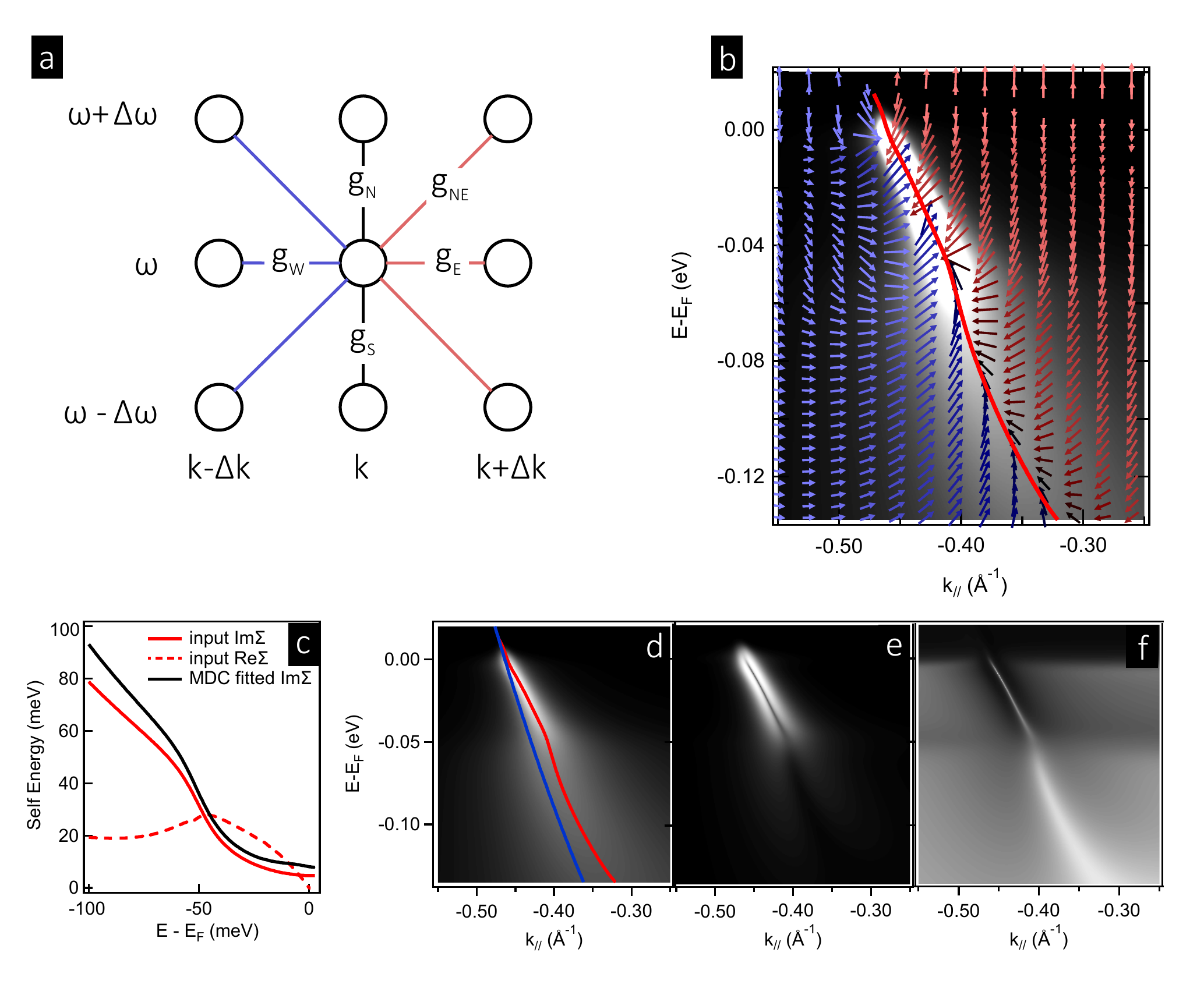}
 \caption{\label{Fig1} (a) The definition of 8 partial gradient components on a 3$\times$3 patch of pixels. (b) gradient vector field for a simulated electron energy-momentum dispersion at 15K. (c) the real and imaginary part of the electron self energy $\Sigma(k,\omega)$ implemented in the simulation. The two quantities satisfy the Kramers-Kronig relation. (d) the simulated spectrum with the bare dispersion (blue) and MDC fitted dispersion (red) on top. (e) the gradient modulus $||G(k,\omega)||$ map. (f) the renormalized gradient modulus map $I(k,\omega)/||G(k,\omega)||$.
 }
 \end{figure*}

The minimum gradient method incarnates the simple spirit of this first order derivative approach. To mitigate the measurement noise in practice, we introduce the \lq redundant gradient vector\rq~$G = (g_N,g_{NE},g_E,g_{SE},g_S,g_{SW},g_W,g_{NW})$, where the redundancy comes from the 8 partial gradient components along the cardinal and intercardinal directions. Fig.~\ref{Fig1}(a) zooms into a 3$\times$3 patch of pixels in the data plot. If we separate the differentiable part and noise part of $I_0(k,\omega)$ into $\widetilde{I_0}(k,\omega)$ and $g_n(\sigma)$, the gradient vector's components $g_N$ and $g_{NE}$ can be defined as:
\begin{equation}
\begin{split}
g_N(k,\omega) & = \frac{I_0(k,\omega + \Delta \omega) - I_0(k,\omega)}{\Delta \omega} \\
& = \partial_\omega \widetilde{I_0}(k,\omega)|_{\omega_0^+} + g_n\big(\frac{2}{\Delta\omega}\sigma\big)\\
& \\
g_{NE}(k,\omega) & = \frac{I_0(k + \Delta k,\omega + \Delta \omega) - I_0(k,\omega)}{\sqrt{(\Delta k)^2 + (\Delta \omega)^2}} \\
& = \partial_{(k+\omega)}\widetilde{I_0}(k,\omega)|_{(k_0+\omega_0)^+} \\
& + g_n\big(\frac{2\sigma}{\sqrt{(\Delta k)^2 + (\Delta \omega)^2}}\big)
\label{eq:9}
\end{split}
\tag{9}
\end{equation}

In the infinitesimal step limit $(\Delta k)^2 + (\Delta \omega)^2 \rightarrow 0$, the discrete 2D map of $\widetilde{I_0}(k,\omega)$ becomes differentiable thus satisfying:
\begin{equation}
\begin{split}
\widetilde{g}_N(k,\omega) & = \partial_{\omega^+} \widetilde{I_0}(k,\omega) = \partial_{\omega} \widetilde{I_0}(k,\omega)\\
& = \partial_{\omega^-} \widetilde{I_0}(k,\omega) = \widetilde{g}_S(k,\omega) \\
& \\
\widetilde{g}_{NE}(k,\omega) & = \bigg(\frac{\Delta k \partial_k + \Delta \omega \partial_\omega}{\sqrt{(\Delta k)^2 + (\Delta \omega)^2}} \bigg)\widetilde{I_0}(k,\omega) \\
& = \frac{\partial_k + \xi\partial_\omega}{\sqrt{1 + \xi^2}}\widetilde{I_0}(k,\omega)
\label{eq:10}
\end{split}
\tag{10}
\end{equation}

Here in Eq.~(\ref{eq:7}) the dimensionless ratio $\xi = \frac{\Delta \omega}{\Delta k}$ is a predetermined quantity from the experimental setup. For simplicity of discussion, we set $\xi = 1$, which represents numerically equal energy and momentum step size in the following discussion. In this case, the modulus of the redundant gradient vector is:
\begin{equation}
\begin{split}
||G|| = &(g_N^2+g_{E}^2+g_S^2+g_{W}^2\\
&+g_{NE}^2+g_{SE}^2+g_{SW}^2+g_{NW}^2)^{\frac{1}{2}} \\
= &2\sqrt{(\partial_k I_0(k,\omega))^2+(\partial_\omega I_0(k,\omega))^2}
\label{eq:11}
\end{split}
\tag{11}
\end{equation}

which is nothing but twice that of the 2-component gradient vector. This means that for \lq intrinsic signal\rq~ there are only 2 degrees of freedom among all the 8 components such that they \lq coherently\rq~ add up. In the mean time, this 8-direction average also enforces robustness when dealing with either very flat or very steep dispersion, because in this case the differential angle can at most be 22.5\textdegree away from the normal of the dispersion slope. Moreover, due to the independent distribution of the noise $g_n(\sigma)$ in $(k,\omega)$ plane, Eqn.~(\ref{eq:11}) effectively performs a spacial average of all 8 independently distributed gaussian noise. To get a rough estimate of the noise level in $||G(k,\omega)||$, we can approximate the total noise by looking at a completely noise dominated scenario, where $g_N \gg \widetilde{I_0}$. In this case, with $\xi = 1$ for simplicity, the total noise will be:
\begin{equation}
\begin{split}
||G_n|| \sim &(g_{n_{N}}^2+g_{n_{E}}^2+g_{n_{S}}^2+g_{n_{W}}^2\\
& + g_{n_{NE}}^2 + g_{n_{SE}}^2+ g_{n_{SW}}^2+ g_{n_{NW}}^2)^{\frac{1}{2}}\\
&
\label{eq:12}
\end{split}
\tag{12}
\end{equation}
where $g_{n_{N,E,S,W}} \sim g_n\big(\frac{2}{\Delta\omega}\sigma\big)$, and $g_{n_{NE,SE,SW,NW}} \sim g_n\big(\frac{\sqrt{2}}{\Delta\omega}\sigma\big)$. The modulus of the noise's gradient vector $||G_n||$ follows a probability distribution very close to the 8-degree-of-freedom chi-squared distribution. But due to the inequivalence of noise variances along cardinal and intercardional directions, numerical calculation is needed to reveal the combined variance to be $\sim 1.26 \frac{\sigma}{\Delta\omega}$. The variance of $||G_n||^{-1}$ for the inverted modulus map becomes $\sim 0.07 \big(\frac{\sigma}{\Delta\omega}\big)^{-1}$.


 \begin{figure*}
 \includegraphics[width=0.9\textwidth]{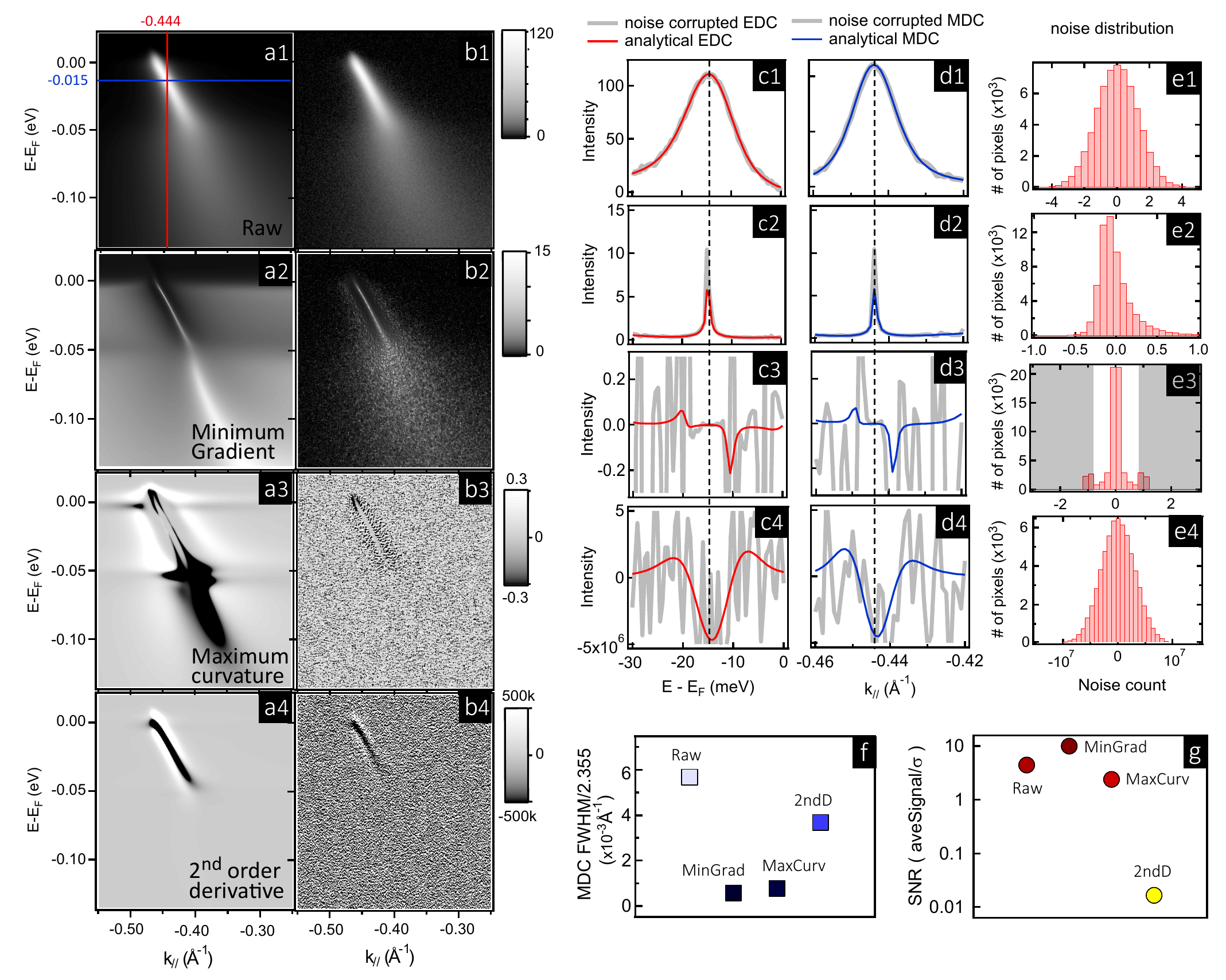}
 \caption{\label{Fig2} (a) The noise-free simulated raw spectra (a1), treated with minimum gradient method (a2), maximum curvature method (a3) and second derivative method (a4) respectively. (b1-b4) Same procedure preformed on Gaussian noise corrupted spectra. The color scales are kept the same between (a) and (b). (c1-c4) EDC plots at $\omega$ = -0.015eV for the simulated spectra (red) and the noise corrupted spectra (grey). (c1-c4) MDC plots at $k$ = -0.444 \AA$^{-1}$ for the simulated spectra (red) and the noise corrupted spectra (grey). (e1-e4) noise distribution profile in plot (b1-b4). (f) the Gaussian fitted standard deviation $\sigma$ (FWHM/2.355) for the noise-free MDC's in (d1-d4). (g) the signal to noise ratio in spectra (b1-b4).
 }
 \end{figure*}

To discuss signal to noise ratio, the signal enhancement as a function of the orders of derivatives also needs to be quantified. In real experiments, the noise propagation is sensitive to the inter-pixel step size $\delta$, where the higher order derivative methods completely rely on the differentiation to increase the signal strength and filter out slow varying background. While there is no universal functional form that captures all the physical quantities being measured, in most cases the signal of interest does not vary drastically on pixel-to-pixel level. In general, as most high pass filters will do, higher order derivative based methods tend to enhance existing sharp features more, and wash out slow varying signals amid drastically enhanced noise. If the derivative is overdone, the sharpened signal will merge with the enhanced noise in the image frequency domain, which practically makes it impossible to identify features. We find the minimum gradient method a good balance between the two aforementioned aspects.

Fig.~\ref{Fig1}(b) plots the simulated spectrum from a tightbinding bareband (blue line in Fig.~\ref{Fig1}(d)) and the self energies constructed in Fig.~\ref{Fig1}(c), with 5~meV by 0.005 \AA$^{-1}$ instrument resolution convolved. The gradient vector field is overlaid on top, where in this case the (vector direction) sign flipping line provides a precise measure of spectral maxima (red line in Fig.~\ref{Fig1}(b) and (d)). The $||G||$ map is shown in Fig.~\ref{Fig1}(e), where the minima of gradient modulus $||G||$ (dark line) traces the maxima of the original spectrum. To separate the slow-varying background from the dispersion locus in $||G||$ map, Fig.~\ref{Fig1}(f) plots the inverse map $||G||^{-1}$ multiplied by the raw spectral intensity $I_0$, making the plot dimensionless in the meantime. It not only shows the strong 50~meV main kink structure and the weak 15~meV sub-kink structure generated by the \emph{ad-hoc} self energy, but also maintains the intensity contrast for dispersion higher than 50~meV, which rarely survives after higher order derivatives due to the feature's low frequency nature.

\subsection{\label{sec:level2}Comparison with Existing Methods}

To check the algorithms' robustness in noise corrupted spectrum, we apply Gaussian noise ($\sigma$ $\sim$ 3\% of maximum spectral signal intensity) to the pristine simulated spectrum and compare the before-and-after performance for the (1) raw spectrum (2) minimum gradient method (3) maximum curvature method (4) second derivative method. Fig.~\ref{Fig2}(a) and (b) show the noise-free and noise-corrupted spectra respectively. The maximum curvature method adopted here uses $C_x = C_y = 1$ per definition in Ref[8].

For the pristine spectra, second derivative method enhances the intense feature ($\omega$ \textgreater -0.05~eV) significantly, but it fails to capture the dispersion with weaker intensity below $\omega \sim$ -0.05~eV. Additionally, due to the strong intensity modulation along $\omega$ direction from the Fermi-Dirac distribution $FD(\omega)$ (Eq.~\ref{eq:4}), an artificial back-bending feature is generated close to $\omega=0$, which could potentially confuse the physical interpretation in energy gap related discussion. In contrast, both minimum gradient method (Fig.~\ref{Fig2}(a2)) and maximum curvature method (Fig.~\ref{Fig2}(a3)) are able to produce sharpened dispersions over the whole energy range with reduced sensitivity to Fermi-Dirac distribution. They are also successful in revealing the dispersion anomalies, although the maximum curvature method does generate ~\lq ripples\rq~ around where it should be a single feature.

However, the three methods show very different noise resilience when the raw data contains a small amount of noise. The signal processed with the maximum curvature method (Fig.~\ref{Fig2}(b3)) and second derivative method (Fig.~\ref{Fig2}(b4)) is immersed in significantly amplified noise due to the high order derivative's nature. The dispersion is only trackable within -0.05~eV~\textless~$\omega$~\textless~0 as a result. On the other hand, the minimum gradient method demonstrates good noise resilience as discussed quantitatively before, and the dispersion anomaly is still well identifiable in Fig.~\ref{Fig2}(b2).
 
To compare more quantitatively, we plot the energy and momentum distribution curves at -0.444~\AA$^{-1}$ (EDC, Fig.~\ref{Fig2}(c)) and -0.015~eV (MDC, Fig.~\ref{Fig2}(d)) for the raw and the processed spectra. For comparability, we keep the plots for pristine and noise corrupted data at the same color scale. The intensity plot range in the EDC/MDC plots are kept the same with that of Fig.~\ref{Fig2}(a) and (b). By comparing the noisy data (cuts from Fig.~\ref{Fig2}(b), grey) and the pristine data (cuts from Fig.~\ref{Fig2}(a), red for EDC and blue for MDC), it can be observed that the increase in the noise level for maximum curvature method (Fig.~\ref{Fig2}(c3)(d3)) and second derivative method (Fig.~\ref{Fig2}(c4)(d4)) is significant (red and blue lines).

By taking the difference between Fig.~\ref{Fig2}(a) and (b), the noise map can be derived. The noise profile of the processed images are plotted in the histograms in Fig.~\ref{Fig2}(e). From Eq.~(\ref{eq:8}), the standard deviation of the noise in Fig.~\ref{Fig2}(e4) is expected to go up by $1.6\times10^7$ times from Fig.~\ref{Fig2}(e1) with $\Delta\omega \sim ~0.0005~eV$, which is consistent with the plot. The noise profile for maximum curvature method has a sharp distribution close to zero whose FWHM is smaller than 0.2, but there is a significant amount of far-outlying noise population which we capped at $\pm$1. This long tail of large valued noise mainly comes from the division process in the calculation of local curvatures, and it can bring in serious interference for the identification of the true signal.

Fig.~\ref{Fig2}(f) lists the Gaussian fitted MDC widths from the noise-free MDC's in Fig.~\ref{Fig2}(d). It is clear that the peak sharpening effect (reduction in FWHM) is strongest in minimum gradient method and maximum curvature method. While the noise distributions for minimum gradient and maximum curvature methods are not Gaussian, the experimental standard deviation can be numerically calculated. If we define the average of the absolute intensity in Fig.~\ref{Fig2}(b) as a consistent measure of signal strength, we can plot and compare the signal to noise ratio for each method (Fig.~\ref{Fig2}(g)). It should be noted that even though the overall signal to noise ratio of maximum curvature method and second derivative method may have deteriorated from that of the noise corrupted raw data, they still sharpen the dispersive feature locally.

 \begin{figure*}[!t]
 \includegraphics[width=0.9\textwidth]{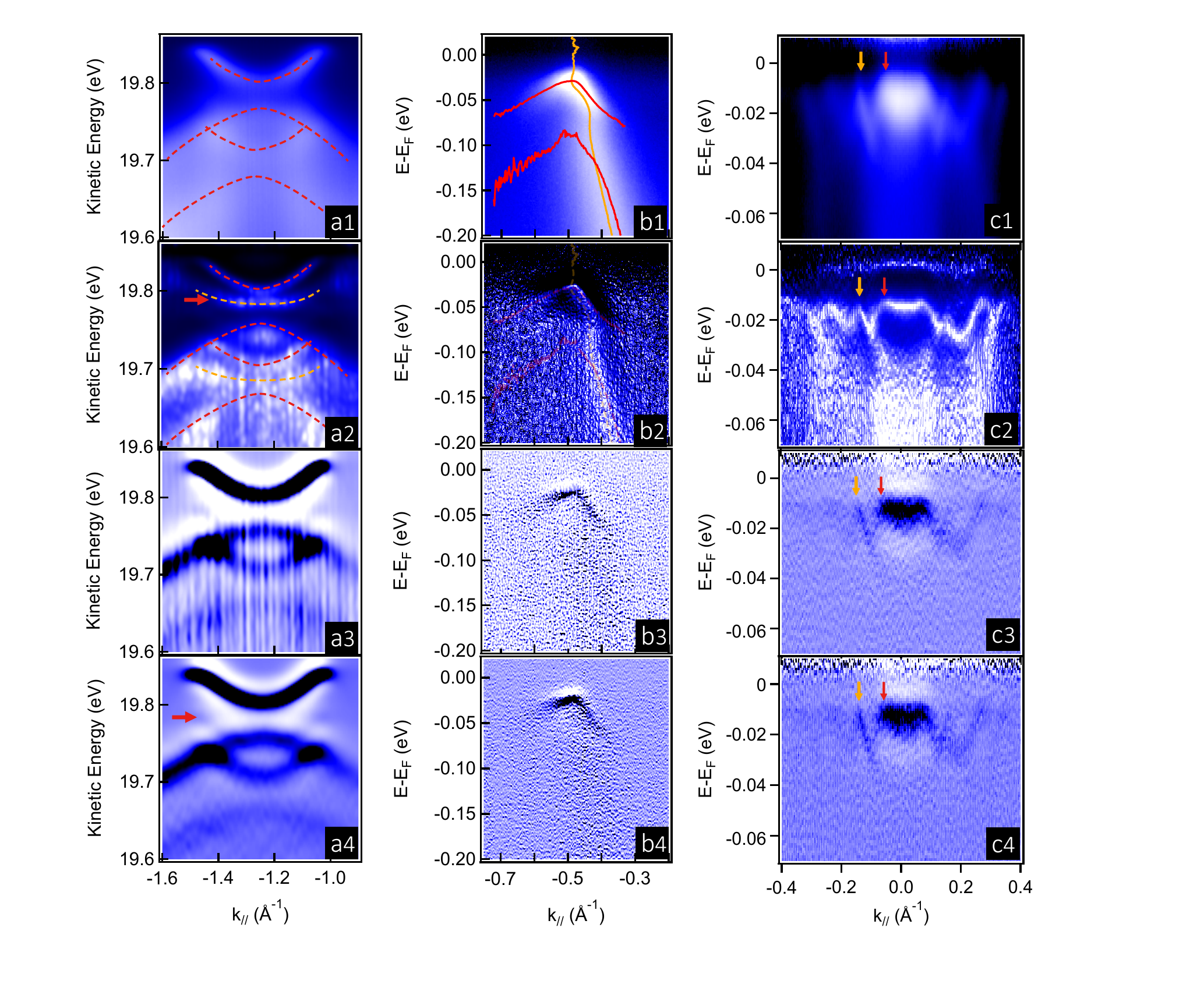}
 \caption{\label{Fig3} (a) The experimental M pocket dispersion acquired for monolayer FeSe film grown on Nb:SrTiO$_3$ at 10K, treated with minimum gradient method (a2), maximum curvature method (a3) and second derivative method (a4) respectively after smoothing. The red arrows indicate the previously missing $d_{xy}$ band. The red dotted lines are guide to the eye for the 1$^{st}$ order phonon shake-off bands.\cite{JJ2014} (b1-b4) The experimental off-nodal dispersion acquired for overdoped Bi2212 single crystal ($T_c = 80K$) at 10K and the subsequent treatments without any smoothing. EDC (red) and MDC (orange) fittings are overlayed on top for comparison. (c1-c4) The experimental antinodal (zone boundary) dispersion acquired for overdoped Bi2212 single crystal ($T_c = 50K$) at 24K and the subsequent treatments without any smoothing. The red and pink arrows indicate the normal state antibonding and bonding band Fermi crossing momenta respectively.
 }
 \end{figure*}

 \begin{figure*}[!t]
 \includegraphics[width=0.9\textwidth]{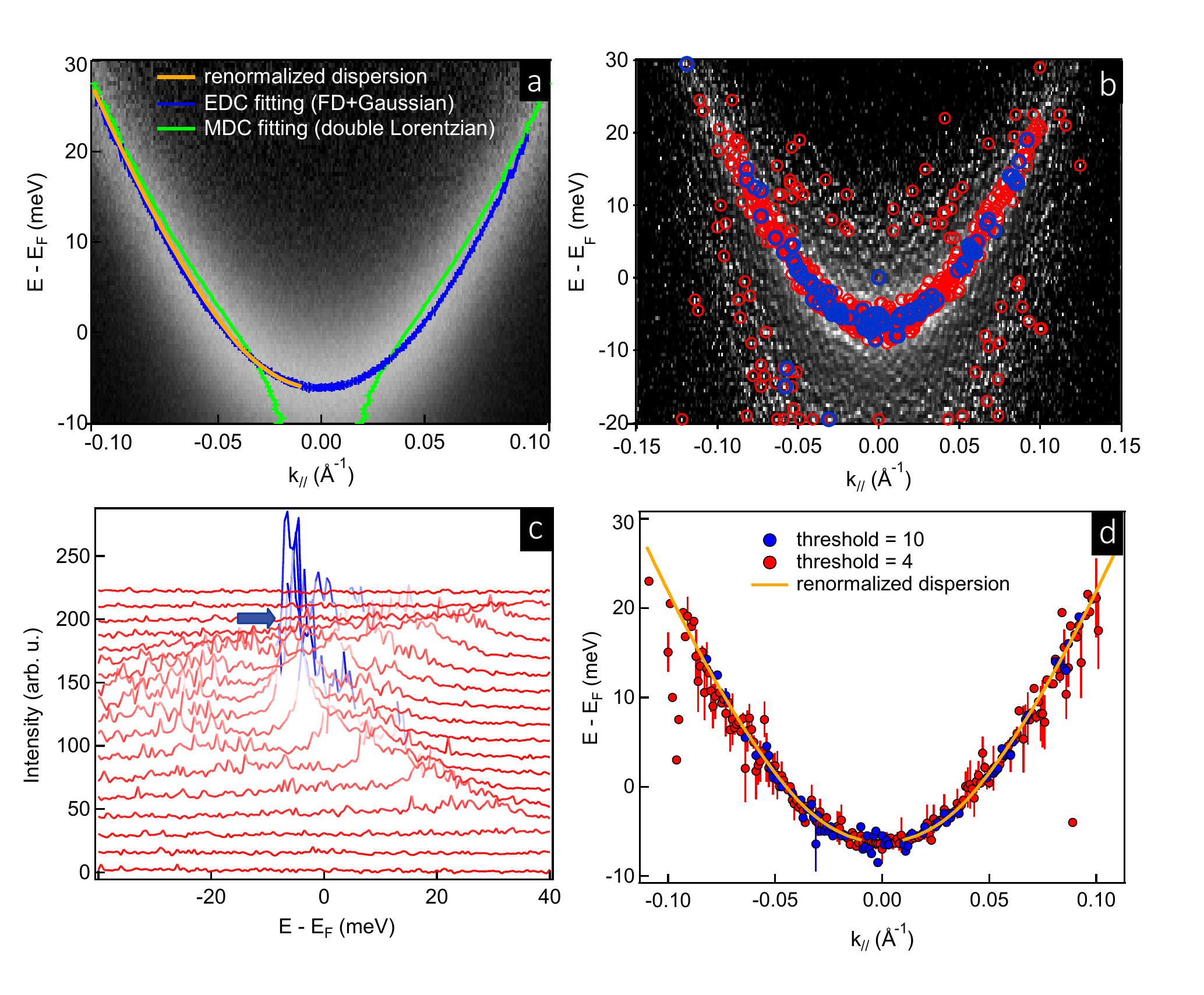}
 \caption{\label{Fig4} (a) Simulated antinodal spectrum with noise corruption for overdoped Bi2212 ($T_c = 50K$) single crystal based on tightbinding parameter fitted from Fig.~(\ref{Fig3}(c1)). (b) intensity renormalized minimum gradient map of Fermi-Drac divided spectrum. Red and blue circles represent domains of definition at threshold values of 4 and 10 respectively. (c) EDC's of (b), with blue indicating the points above threshold value. (d) Weighed energy-momentum dispersion calculated on the domain of definition from (b) for different threshold values.
 }
 \end{figure*}
 
\subsection{\label{sec:level2}Applications to ARPES spectrum}

To test and compare the performance among the aforementioned methods in real systems, we carry out the analyses in ARPES data on monolayer FeSe superconducting film (Fig.~\ref{Fig3}(a)) and bi-layer cuprate superconductors' near nodal (Fig.~\ref{Fig3}(b)) and antinodal (Fig.~\ref{Fig3}(c)) region. It is demonstrated that the new method is effective both in bringing out faint but physically important features and in noise suppression. The dimensionless nature of plotting $I_0/||G||$ also gives the processed spectrum a relatively system-independent intensity scale, usually between 0 and 10.

Fig.~\ref{Fig3}(a1) shows the raw spectrum of a $\Gamma M$ cut in monolayer FeSe/STO film at $T = 10K$. Established studies identify a set of ~\lq shadow bands\rq~ (red dashed lines) at about 0.1~eV below the $0^{th}$ order bands. However, a third band ($d_{xy}$ character) which is normally weaker than the other two has not been identified. Meanwhile, from second derivative process (Fig.~\ref{Fig3}(a4)), a faint shadow was noticed by Lee et al.\cite{JJ2014,he2016} but with unexplained origin. With the minimum gradient method (Fig.~\ref{Fig3}(a2)), the missing $d_{xy}$ band (orange dashed line) shows up both in the original and the shadow bands after smoothing. The fact that Maximum curvature method performs similarly (Fig.~\ref{Fig3}(a3)) to second derivative method in this case.

Fig.~\ref{Fig3}(b1) shows the raw spectrum of an off-nodal cut in a hole overdoped Pb-Bi$_2$Sr$_2$CaCu$_2$O$_8$ single crystal ($T_c = 80K$) at $T = 10K$. With EDC (red line) and MDC (orange line) fitting, the band dispersion can be extracted with considerable discrepancy. At superconducting gap edge, MDC fitting completely fails to capture the band backbending behavior due to flat band top; and near/below the dispersion anomaly $\sim -0.07~eV$, EDC fitting becomes inaccurate due to the steepness of the dispersion. Minimum gradient method (Fig.~\ref{Fig3}(b2)) has the advantage of differentiation direction insensitivity, thus successfully combines the advantage of EDC and MDC fitting at different parts of the dispersion. Maximum curvature method and second derivative method (Fig.~\ref{Fig3}(b3)(b4)) emphasize on the intense features in the raw spectrum, but expectedly wash out features below the dispersion anomaly.

Fig.~\ref{Fig3}(c1) shows the raw spectrum of an antinodal cut in an extremely hole overdoped Bi$_2$Sr$_2$CaCu$_2$O$_8$ single crystal ($T_c = 50K$) at $T = 24K$. The system hosts 3 superstructure-generated replica bands parallel in this momentum, which makes it difficult to identify each set of the bands. Maximum curvature method and second derivative method (Fig.~\ref{Fig3}(c3)(c4)) only picks out the first set of bonding (orange arrow) and antibonding (red arrow) bands. The minimum gradient method (Fig.~\ref{Fig3}(c2)) is able to sharpen the dispersion, resolve the superconducting gap backbending and identify more higher order bands from the superstructure.

\subsection{\label{sec:level2}Extraction of 1D Dispersion}
 
In order to facilitate further quantitative analysis, we propose a way to extract the 1D dispersion curve from the 2D image map. Fig.~\ref{Fig4}(a) shows the intensity plot of a simulated dispersion, with MDC fitting (green), EDC fitting (blue) and input dispersion (orange, only left half is shown) overlayed on top. The discrepancy between MDC and EDC fitting extracted dispersion is clearly reflected.

Fig.~\ref{Fig4}(b) is the minimum gradient method sharpened intensity map and its EDC's are plotted in Fig.~\ref{Fig4}(c). By introducing a \lq threshold\rq~ parameter (blue arrow in Fig.~\ref{Fig4}(c)), all the pixels with intensity above the threshold (blue part of EDC's in Fig.~\ref{Fig4}(c)) composite a new, downsized domain of definition (blue circles in Fig.~\ref{Fig4}(b)). At last, for each momentum ($k$), a weighted energy average ($\omega$) is calculated on the downsized new domain with the weighting function being the raw image intensity. This will yield a 1D curve of $\omega(k)$, which is represented by the blue dispersion Fig.~\ref{Fig4}(d). It agrees with the input dispersion consistently well both at the flat band bottom and the rapidly dispersing region.

To test the stability of the free threshold parameter, we intentionally loosen the domain selection requirement by lowering the threshold from 10 to 4. As indicated in Fig.~\ref{Fig4}(b), the domain for weighted average increases from the original blue circles (threshold = 10) to all the red circles (threshold = 4). Nonetheless, the weighted position still closely follows the input dispersion in the lower threshold's case over the entire fitting range with few than 5\% outliers. The stability comes from the sharp contrast and relatively low noise level in the minimum gradient map, which is crucial in the domain selection process.

 \begin{figure*}[!t]
 \includegraphics[width=0.9\textwidth]{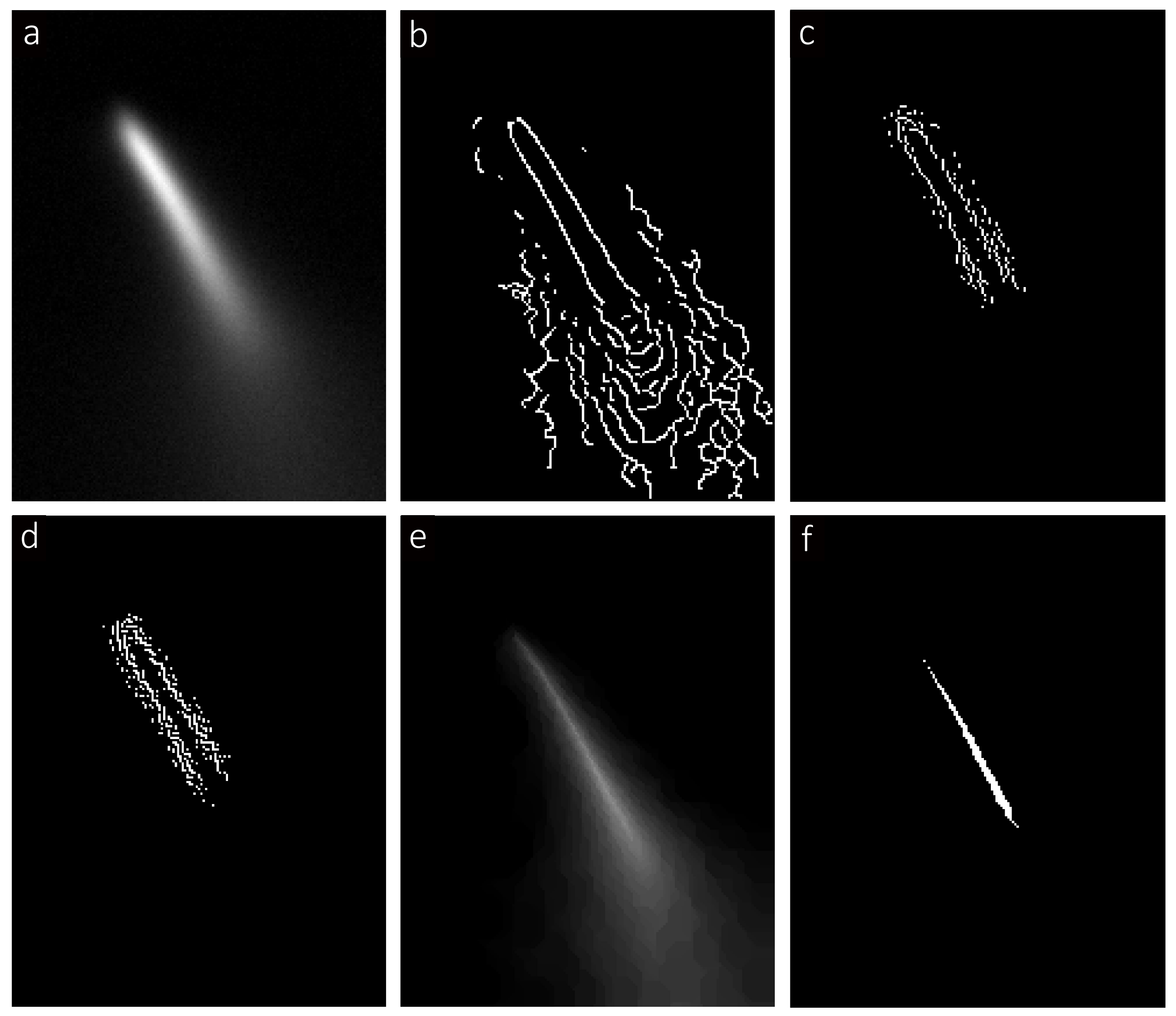}
 \caption{\label{Fig6} (a) Noise corrupted simulated nodal dispersion as used in Fig.~\ref{Fig2}(b1). (b)(c)(d) Edge detection with Canny filter, Sobel filter and Roberts filter. (e) Eroded dispersion with disk size of 8 pixels. (f) Binarized image skeleton (dispersion) from (e), with binarization threshold set at 0.6.
 }
 \end{figure*}

\section{\label{sec:level1}DISCUSSION}

One caveat for the minimum gradient algorithm is the potential false detection of intensity minima. In the process of gradient vector modulus $||G||$ minimization, intensity valleys are indistinguishable from intensity ridges. This issue is in part alleviated by weighing the reciprocal of the gradient vector modulus with the raw spectral intensity $I_0$. This can be more aggressively resolved by multiplying a weighing factor that scales with magnitude of the local Hessian determinant, such that local minima are further suppressed. However, this process would require second order derivatives, and will inevitably lead to similarly increased noise level as other higher order derivative based methods would encounter.

Put in a broader scope, the minimum gradient method here is similar in spirit to the Kirsch operator ~\cite{KIRSCH1971}, but utilized to locate the intensity maxima instead of edges. Classic edge detection algorithms like Canny and Sobel filter typically yield the contour of the most prominent peak feature in the spectra ~\cite{canny1986}, but remains susceptible to noise corruption. Morphological image erosion can also be used to extract the image skeleton, but for quantitative analysis purpose, it would require more comprehensive discussion on structure element selection and relevant operations in the scale space ~\cite{tw2016}.

\section{\label{sec:level1}SUMMARY}

In this manuscript, we carry out comprehensive comparison among popular 2D image sharpening algorithms in physical research, from the perspectives of both the image's primary domain and frequency domain. A minimum gradient based method is proposed to balance the signal to noise performance and the image sharpness. This method is shown to be robust against noise, and is known to moderately preserve weak intensity feature. It is also mathematically exact in determining the intensity maxima position. These traits are particularly helpful to quickly reveal subtle features in real-time spectra acquisition, which usually requires involved thus time-consuming data processing for more quantitative analysis.

\section{\label{sec:level1}ACKNOWLEDGEMENT}
The authors want to thank Xiujun Zhu and Jue Fan for discussion on the propagation of noise distribution and Makoto Hashimoto for the application to ARPES spectra. James Lee provided the photoemission data on FeSe monolayer films. The study is supported by the Department of Energy, Office of Basic Energy Sciences, Division of Materials Sciences and Engineering.

%


\end{document}